\documentclass[showpacs,pre]{revtex4}
\usepackage{graphicx}

\begin{document}

\title{Two-dimensional solitons in the Gross-Pitaevskii equation with
spatially modulated nonlinearity}
\author{Hidetsugu Sakaguchi$^{1}$ and Boris A. Malomed$^{2}$}

\address{$^{1}$Department of Applied Science for Electronics and Materials,\\
Interdisciplinary Graduate School of Engineering Sciences,\\
Kyushu University, Kasuga, Fukuoka 816-8580, Japan\\
$^{2}$Department of Interdisciplinary Studies,\\
School of Electrical Engineering, Faculty of Engineering,\\
Tel Aviv University, Tel Aviv 69978, Israel}

\begin{abstract}
We introduce a dynamical model of a Bose-Einstein condensate based
on the 2D Gross-Pitaevskii equation, in which the nonlinear
coefficient is a function of radius. The model describes a
situation with spatial modulation of the the negative atomic
scattering length, via the Feshbach resonance controlled by a
properly shaped magnetic of optical field. We focus on the
configuration with the nonlinear coefficient different from zero
in a circle or annulus, including the case of a narrow ring.
Two-dimensional solitons are found in a numerical form, and also
by means of a variational approximation; for an infinitely narrow
ring, the soliton is found in an exact form. A stability region
for the axisymmetric solitons is identified by means of numerical
and analytical methods. In particular, if the nonlinearity is
supported on the annulus, the upper stability border is determined
by azimuthal perturbations; the stability region disappears if the
ratio of the inner and outer radii of the annulus exceeds a
critical value $\approx 0.47$. The model gives rise to
bistability, as the stationary solitons coexist with stable
breathers, whose stability region extends to higher values of the
norm than that of the static solitons. The collapse threshold
strongly increases with the radius of the inner hole of the
annulus. Vortex solitons are found too, but they are unstable.
\end{abstract}

\maketitle

\section{Introduction}

Matter-wave solitons have been created in Bose-Einstein
condensates (BECs) in various effectively one-dimensional (1D)
settings. First, these were dark solitons in repulsive condensates
\cite{dark}. Then, bright solitons were created in an attractive
BEC (lithium) \cite{bright}. This was followed by the making of
gap solitons in a repulsive rubidium condensate loaded in a
periodic potential, which was induced by the optical lattice (OL),
i.e., interference pattern between two laser beams illuminating
the medium \cite{Oberthaler}.

A challenge to the experiment is creation of 2D matter-wave
solitons. A natural problem in this case is the trend of solitons
in multidimensional attractive condensates to be unstable because
of the possibility of collapse in this setting \cite{collapse}. In
theoretical works, several approaches were proposed to stabilize
2D solitons. One of them relies on the use of a full
two-dimensional OL \cite{2D-OL}, or its low-dimensional (quasi-1D)
counterpart \cite{Q1D-OL}, which can stabilize fundamental
solitons. In addition, 2D lattices lend stability to vortical
solitons \cite{2D-OL}, including higher-order vortices, and
``super-vortex" complexes \cite{we}; the latter are built as
circular chains of compact vortices, with global vorticity imposed
on top of the chain. Another theoretically elaborated approach
relies upon the use of a nonlocal anisotropic nonlinearity induced
by the long-range interactions between atoms with a magnetic
momentum (chromium), polarized by an external field
\cite{dipolar}.

An alternative mechanism proposed for the stabilization of 2D matter-wave
solitons is based on the Feshbach resonance (FR), which makes it possible to
control the value of the scattering length, i.e., as a matter of fact, an
effective nonlinear coefficient in the corresponding Gross-Pitaevskii
equation (GPE), by means of an external magnetic field \cite{FR}. Moreover,
the FR may switch the sign of the nonlinearity (in particular, the
FR-induced switch from repulsion to weak attraction was instrumental to the
creation of bright solitons in lithium \cite{bright}). Application of a
low-frequency ac magnetic field may provide for periodic alternation of the
nonlinearity sign in the GPE via the FR. It was predicted that the FR
technique based on the ac field gives rise to novel states in the 1D
geometry \cite{FRM}, and can stabilize 2D solitons, even in the absence of
the external trap \cite{2Dstabilization}. The same technique, if applied in
combination with a quasi-1D OL potential, may also stabilize matter-wave
solitons in the 3D geometry \cite{Michal}.

It has been predicted \cite{Optical-theory}, and demonstrated in
experiment \cite{Optical-experiment}, that the FR can also be
induced by a properly tuned optical field. Then, illuminating the
condensate by two counterpropagating coherent laser beams, one can
build an OL that will provide for periodic modulation of the
nonlinearity coefficient along the respective spatial coordinate.
Solitons in the corresponding one-dimensional GPE with the
\emph{nonlinear} OL were recently investigated in Ref. \cite{1d},
where stability regions for static solitons and breathers were
found (motion of free solitons in the same model was recently
studied in Ref. \cite{Fatkhulla}, and rigorous proofs concerning
the stability of static solutions in this setting were reported in
work \cite{MIW}). The soliton dynamics in the 1D model with other
configurations of the spatial modulation of the nonlinearity
coefficient was studied in Refs. \cite{DimPan} (unlike Ref.
\cite{1d}, the nonlinearity coefficient did not change its sign in
the models considered in the latter works).

Static spatial modulation of the nonlinearity through the FR,
controlled by the properly shaped magnetic or optical field, may
be tried as another means for the stabilization of 2D solitons,
which is the subject of the present work. A natural form of am
axisymmetric OL in the 2D geometry corresponds to the Bessel beam,
i.e., a nondiffracting light signal in a bulk linear medium. In
the case when the Bessel beam creates an effective linear
potential in the equation of the GPE type with self-attraction, it
has been shown that the corresponding radial lattice can readily
stabilize various types of 2D solitons \cite{Barcelona}. However,
our results show that, within a broad parameter region that we
were able to explore, stabilization of 2D solitons by means of a
\emph{nonlinear} Bessel lattice, i.e., within the framework of the
GPE whose nonlinear coefficient is $g(r)=g_{0}J_{n}(ar)$, where
$r$ is the radial coordinate, $g_{0}$ and $a$ are constants, and
$J_{n}$ is the Bessel function with $n=0,1,...$, appears to be
\emph{impossible} -- stationary axisymmetric soliton solutions can
be easily constructed, but in simulations they all suffer either
decay or collapse.

Nevertheless, in this work we demonstrate that a simpler shape of the radial
modulation of the nonlinearity, in which it takes a constant value,
corresponding to self-attraction, inside a finite circle or annulus, and is
zero (or corresponds to self-repulsion) outside this region, is able to
stabilize axisymmetric 2D solitons. In addition to that, we will demonstrate
that the model gives rise to \emph{bistability}: the stationary solitons
coexist with stable breathers, that feature persistent oscillations in the
radial direction. In fact, the stability region of the breathers is larger
than that of the static solitons, extending to higher values of the norm
(number of atoms in the BEC).

It should be said that, in the case of the nonlinearity controlled
by the optical beam through the FR mechanism, the beam with the
cross section in the form of a circle or annulus is not
divergence-free, unlike its Bessel-shaped counterpart. However,
this circumstance does not impede the physical realization of the
model, as an effectively 2D condensate can be easily trapped
between two blue-detuned light sheets, which strongly repel the
atoms, as demonstrated in the experiment \cite{pancake}. The
thickness of the corresponding ``pancake" is a few microns, while
its diameter is measured in hundreds of microns (at least), hence
the diffraction of the light beam within this range is completely
negligible.

The paper is organized as follows. In Section 2, we give the formulation of
the model, and present numerical and analytical solutions for static
solitons. The analytical part includes a variational approximation for the
solutions in the general case, an exact solution for solitons supported by
an infinitely narrow annulus carrying the nonlinearity, and predictions for
the stability against radial perturbations, based on the Vakhitov-Kolokolov
(VK) \cite{VK,Berge'} criterion. The stability threshold for azimuthal
perturbations is determined by a solution of the corresponding eigenvalue
problem. An inference is that stability borders in the model with the
nonlinearity supported on the circle are completely determined by radial
perturbations, while in the annular model the upper stability border (in
terms of the soliton's norm) is controlled by azimuthal perturbations. No
stable solitons are possible if the annulus is relatively narrow, with the
ratio of inner and outer radii exceeding a critical value $\approx 0.47$. In
Section 3, we summarize results of direct numerical simulations of the
stability of fundamental stationary solutions, which precisely confirm the
existence of a well-defined stability region of the 2D solitons in the
model's parameter space, predicted in Section 2. The bistability
(coexistence of the stable stationary solitons and breathers) and the
extended stability region for the breathers are also reported in Section 3.
In Section 4, we briefly consider solitons with intrinsic vorticity, and
conclude that all the vortices are unstable (the vortex splits in two
fundamental solitons, each one then collapsing intrinsically). The paper is
concluded by Section 5.

\section{Stationary solitons}

\subsection{The model and numerical solutions}

The GPE for the single-atom wave function $\psi $ in the normalized form is
\begin{equation}
i\frac{\partial \psi }{\partial t}=-\frac{1}{2}\nabla ^{2}\psi -g(r)|\psi
|^{2}\psi ,  \label{GPE}
\end{equation}
with $t$ time, $\nabla ^{2}$ the 2D Laplacian, and the nonlinearity
coefficient shaped, by means of the external magnetic or optical field, as
said above:
\begin{equation}
g(r)=\left\{
\begin{array}{cc}
1, & \rho <r<R, \\
0, & r<\rho ~\mathrm{or}~r>R.\end{array}\right.   \label{g}
\end{equation}
The number of atoms is determined by the norm of the wave function,
\begin{equation}
N=2\pi \int_{0}^{\infty }|\psi (r)|^{2}rdr.  \label{N}
\end{equation}
Using the scaling invariance of Eq. (\ref{GPE}), we set $R=2$, keeping $\rho
$ as a free parameter. Note that the model without the inner orifice, $\rho
=0$, is a universal one, as it contains no parameters.

We also considered a model with the nonlinearity switched to self-repulsion,
i.e., $g(r)<0$, in the regions of $r<\rho $ and $r>R$. However, we focus on
the case with $g=0$ in these regions, as such a case is least favorable for
the existence of solitons, hence it provides for results which are most
relevant to the experimental realization of the scheme.

Stationary solutions for fundamental solitons are looked for as $\psi =\phi
(r)e^{-i\mu t}$, $\ $with a real chemical potential $\mu $, and a real
function $\phi $ obeying the equation
\begin{equation}
2\mu \phi +\phi ^{\prime \prime }+r^{-1}\phi ^{\prime }+2g(r)\phi ^{3}=0
\label{phi}
\end{equation}
(the prime stands for $d/dr$). Equation (\ref{phi}) is to be solved with the
boundary conditions $\phi ^{\prime }(r=0)=0$ and $\phi (r=\infty )=0$ (the
latter one implies that $\mu $ must be negative). The solution was searched
for numerically by selecting the value of $\phi (r=0)$ with which the
boundary condition at $r=\infty $ could be met.

Two examples of the solution are displayed in Fig. 1(a), one for
$\rho =0$, i.e., the configuration with no inner ``hole", and the
other one with the ``hole" corresponding to $\rho =0.5$; in the
latter case, the solution attains a maximum at $r=\rho $, having a
shallow minimum at $r=0$. Families of the soliton solutions are
characterized by dependences $\mu (N)$, which are displayed in
Fig. 1(b) for $\rho =0$ and two nonzero values of $\rho $. These
dependences predict a necessary stability condition as per the VK
criterion \cite{VK}, $dN/d\mu <0$, i.e., parts of the solution
families beneath the turning points in Fig. 1(b) may be stable
(below, the turning point will be denoted as
$N=N_{\mathrm{cr}}^{(\mathrm{lower)}}$). In fact, the stability
region exists due to the fact that the attractive nonlinearity
acts in a finite region of space, $r<R$.
\begin{figure}[tbp]
\includegraphics[height=5cm]{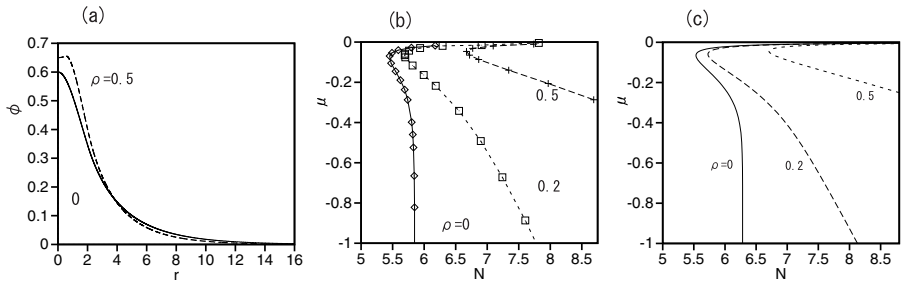}
\caption{(a) Examples of stable soliton solutions with
$\protect\rho =0$, $\protect\mu =-0.0399$, and $N=5.59$ (solid
curve) and $\protect\rho =0.5$, $\protect\mu =-0.0648$, and
$N=6.721$ (dashed curve). (b) Chemical potential $\protect\mu $
vs. norm $N$ for soliton families found numerically with
$\protect\rho =0$, $0.2$, and $0.5$. For $\protect\rho =0$, the
\textit{VK-stable} portion of the solution, i.e., one with
$dN/d\protect\mu <0$, is found in the interval
$N_{\mathrm{cr}}^{(\mathrm{lower)}}\approx
5.449<N<N_{\mathrm{Townes}}\approx 5.85$. (c) $\protect\mu (N)$
curves predicted by the variational approximation for the same
cases, $\protect\rho =0$, $0.2$, and $0.5$.} \label{fig1}
\end{figure}

In the absence of the inner orifice ($\rho =0$), the soliton becomes narrow
as $\mu $ takes large negative values. In this case, the medium seems nearly
uniform for the soliton, hence it approaches the shape of the well-known
\textit{Townes soliton}, which is a universal weakly unstable localized
solution of the 2D nonlinear Schr\"{o}dinger (NLS) equation with the
spatially uniform self-focusing nonlinearity \cite{Berge'}. Accordingly, the
soliton's norm approaches the value $N_{\mathrm{Townes}}\approx 5.85$, which
plays a critical role in the radial dynamics, being equal to the norm of the
Townes soliton.

\subsection{Variational approximation}

The fundamental soliton solutions in the present model can also be obtained
by means of the variational approximation (see a review of the method in
Ref. \cite{Progress}). To this end, we adopt the ansatz
\begin{equation}
\phi =A\exp \left( -\frac{r^{2}}{2w^{2}}\right) ,  \label{ansatz}
\end{equation}
with an amplitude $A$ and width $w$. The substitution of the
ansatz in norm (\ref{N}) and Lagrangian of Eq. (\ref{phi}),\begin{equation}
L=2\pi \int_{0}^{\infty }\left[ 2\mu \phi ^{2}-\left( \frac{d\phi
}{dr}\right) ^{2}+g(r)\phi ^{4}\right] rdr,  \label{L}
\end{equation}
yields $N=\pi A^{2}w^{2}$ (we use this relation to eliminate $A$ in favor of
$N$), and
\[
L=2\mu N-\frac{N}{w^{2}}+\frac{N^{2}}{2\pi w^{2}}\left(
1-e^{-2R^{2}/w^{2}}\right) .
\]Then, the variational equations, $\partial L/\partial N=0$ and $\partial
L/\partial w=0$, predict the following relations between the norm, width and
chemical potential of the soliton,
\begin{eqnarray}
\frac{2\pi }{N} &=&\left( 1-2\frac{\rho ^{2}}{w^{2}}\right) e^{-2\rho
^{2}/w^{2}}-\left( 1-2\frac{R^{2}}{w^{2}}\right) e^{-2R^{2}/w^{2}},
\nonumber \\
\mu w^{2} &=&1-\frac{N}{2\pi \left( e^{-2\rho
^{2}/w^{2}}-e^{-2R^{2}/w^{2}}\right) }.  \label{VA}
\end{eqnarray}
The $\mu (N)$ dependence, predicted by Eq. (\ref{VA}), is shown in Fig. 1(c)
for several values of $\rho $. It is consistent with the numerical results
displayed in Fig.~1(b), although the variational approximation predicts
somewhat larger values of $N$.

\subsection{The narrow-ring model}

The simple ansatz (\ref{ansatz}) cannot predict the shape of the
solution with the local minimum at $r=0$, such as the one shown in
Fig. 1(a) for $\rho \neq 0$. The minimum becomes deeper as the
nonlinearity-supporting annulus narrows, which corresponds to
$\left( R-\rho \right) /R\rightarrow 0$. As a limit form, one can
take the GPE with the $\delta $-functional nonlinearity support,
\begin{equation}
i\frac{\partial \psi }{\partial t}=-\frac{1}{2}\nabla ^{2}\psi -\delta
(r-R)|\psi |^{2}\psi   \label{delta}
\end{equation}
[the coefficient in front of the $\delta $-function is scaled to be $1$, cf.
Eq. (\ref{g})]. By final rescaling, one can again set $R=2$ in Eq.
(\ref{delta}), as was done above in Eq. (\ref{GPE}), so as to cast
Eq. (\ref{delta}) in a parameter-free form. It is relevant to
mention that a BEC configuration in the form of a narrow ring was
recently created in the experiment by means of an accordingly
shaped magnetic trap \cite{ring}.

In the present case, the stationary wave function $\phi (r)$ obeys a linear
equation,
\begin{equation}
\frac{d^{2}\phi }{dr^{2}}+\frac{1}{r}\frac{d\phi }{dr}+2\mu \phi =0,
\label{linear}
\end{equation}
which must be solved separately for $r<R$ and $r>R$. The inner and outer
solutions, one with $\phi ^{\prime }(r=0)=0$ and the other
vanishing at $r\rightarrow \infty $, are to be linked by the
conditions of the continuity of $\phi (r)$ and jump of $\phi
^{\prime }(r)$ at $r=R$, which follows from Eq. (\ref{delta}):
\begin{equation}
\phi ^{\prime }(r=R+0)-\phi ^{\prime }(r=R-0)=-2\left[ \phi (r=R)\right]
^{3}.  \label{r=R}
\end{equation}
Appropriate solutions to Eq. (\ref{linear}) are\begin{equation}
\phi (r)=A\left\{
\begin{array}{cc}
I_{0}\left( \sqrt{-2\mu }r\right) /I_{0}\left( \sqrt{-2\mu }R\right) , & r<R,
\\
K_{0}\left( \sqrt{-2\mu }r\right) /K_{0}\left( \sqrt{-2\mu
}R\right) , & r>R,\end{array}\right.   \label{Bessel}
\end{equation}
where $I_{0}$ and $K_{0}$ are the modified Bessel and Hankel functions, $A$
is a constant, and the continuity of $\phi (r)$ at $r=R$ is
provided automatically. The substitution of expressions
(\ref{Bessel}) in Eq. (\ref{r=R}) yields
\begin{equation}
A^{2}=\sqrt{-\frac{\mu }{2}}\left[ \frac{K_{1}\left( z\right)
}{K_{0}\left( z\right) }+\frac{I_{1}\left( z\right) }{I_{0}\left(
z\right) }\right] |_{z=\sqrt{-2\mu }R}~.  \label{A^2}
\end{equation}
The norm (\ref{N}) of the exact solution given by
Eqs. (\ref{Bessel}) and (\ref{A^2}) can also be calculated in an explicit form:
\begin{equation}
N=\pi \sqrt{-\frac{\mu }{2}}\left[ \frac{K_{1}\left( z\right)
}{K_{0}\left( z\right) }+\frac{I_{1}\left( z\right) }{I_{0}\left(
z\right) }\right] \left[ \frac{K_{1}^{2}\left( z\right)
}{K_{0}^{2}\left( z\right) }-\frac{I_{1}^{2}\left( z\right)
}{I_{0}^{2}\left( z\right) }\right] |_{z=\sqrt{-2\mu }R}~.
\label{NBessel}
\end{equation}
Figures 2(a) and (b) display, respectively, an example of the solution, and
the $\mu (N)$ dependence plotted as per the exact expression
(\ref{NBessel}).
\begin{figure}[tbp]
\includegraphics[height=3.5cm]{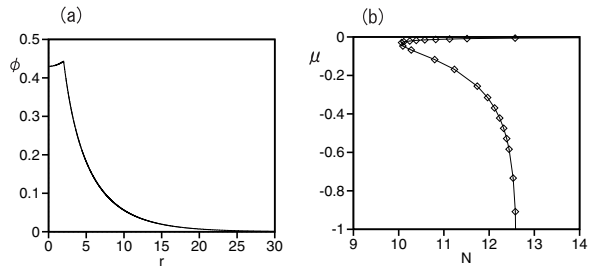}
\caption{(a) An example of the solution of Eq. (\protect\ref{delta}). (b)
The $\protect\mu (N)$ dependence for the $\protect\delta $-functional model,
according to Eq. (\protect\ref{NBessel}).}
\label{fig2}
\end{figure}

\subsection{Stability diagram for stationary solitons}

Figure 2(b) shows the existence of solutions with $dN/d\mu <0$ in the model
with the radial $\delta $-function, which may be stable according to the VK
criterion. However, it can only guarantee the stability against radial
perturbations that do not break the axial symmetry of the solutions. On the
other hand, it is well known that axisymmetric ring-shaped states may be
easily subject to instability against azimuthal perturbations (see, e.g.,
Refs. \cite{azimuthal}).

To study the stability against angular modulations in the general case [with
$g(x)$ taken as per Eq. (\ref{g})], including the $\delta $-functional
limit, as in Eq. (\ref{delta}), we take a perturbed solution as
\begin{equation}
\psi (r,\theta ,t)=e^{-i\mu t}[\phi (r)+\delta \phi _{+}(r)e^{-i\chi
t+im\theta }+\delta \phi _{-}(r)e^{i\chi ^{\ast }t-im\theta }],  \label{pert}
\end{equation}
where $\theta $ is the angular variable, $m$ is an integer perturbation
index, $\chi $ is a perturbation eigenfrequency, with $\ast $
standing for the complex conjugation ($\chi $ may be complex
\cite{azimuthal}), and $\delta \phi _{\pm }(r)$ are components of
the respective eigenfunction. In particular, the instability
threshold may correspond to $\chi =0$, then the eigenfunction has
$\delta \phi _{+}=\delta \phi _{-}\equiv \delta \phi _{0}(r)$, and
the substitution of expression (\ref{pert}) in Eq. (\ref{delta})
and subsequent linearization lead to an equation for the
\textit{zero mode},
\begin{equation}
\left[ \mu +\frac{1}{2}\left(
\frac{d^{2}}{dr^{2}}+\frac{1}{r}\frac{d}{dr}-\frac{m^{2}}{r^{2}}\right)
+3g(r)\left( \phi (r)\right) ^{2}\right] \delta \phi _{0}=0.
\label{delta-phi}
\end{equation}
The instability threshold is achieved when real $\mu $, found as an
eigenvalue of Eq. (\ref{delta-phi}), coincides with the actual
value of the chemical potential of the unperturbed solution $\phi
(r)$. This way, the threshold was identified for the lowest
azimuthal perturbation mode, with $m=1$ (in direct simulations
presented in the next section, instability was observed solely
against the azimuthal modulations with $m=1$).

The result of the analysis is summarized in Fig. 3, in the form of a
stability diagram in the $(\rho ,N)$ parameter plane. The upper dotted
border is the critical curve for the azimuthal instability with $m=1$, found
as described above, while the lower dashed curve is the existence and
stability border for the soliton solutions, which is identified as a set of
turning points of the $\mu (N)$ curves in Fig. 1. Soliton solutions
satisfying the VK criterion, $dN/d\mu <0$, exist above the lower border.
Below the upper border, they are stable against the $m=1$ azimuthal
disturbances, i.e., the solitons are expected to be completely stable
between the two curves. This expectation was verified by direct simulations,
see the next section.

\begin{figure}[tbp]
\includegraphics[height=3.5cm]{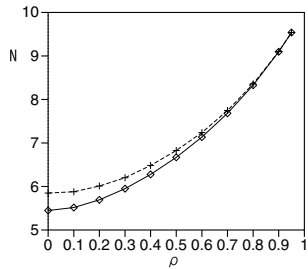}
\caption{The stability diagram for the soliton solutions. In the region
between the two borders, the stationary solitons are stable --
simultaneously according to the VK criterion, i.e., against radial
perturbations (above the lower border), and against azimuthal modulations
(below the upper border). }
\label{fig3}
\end{figure}

Note that the border of the azimuthal instability in Fig. 3 is
located, for $\rho =0$, at a value of $N$ which is identical to
$N_{\mathrm{Townes}}\approx 5.85$, i.e., in the case of $\rho =0$
(no inner orifice), the thresholds for the collapse in the radial
direction, and for the breakup of the axial symmetry in the
azimuthal direction, are identical. The coincidence of the two
thresholds for $\rho =0$ can be explained. Indeed, differentiation
of Eq. (\ref{phi}) in $r$ shows that, for given $\phi (r)$, the
function $\phi ^{\prime }(r)$ solves the following linear
equation:
\begin{equation}
\left[ 2\mu
+\frac{d^{2}}{dr^{2}}+\frac{1}{r}\frac{d}{dr}-\frac{1}{r^{2}}+6g(r)\left(
\phi (r)\right) ^{2}\right] \phi ^{\prime }=-2g^{\prime }(r)\left(
\phi (r)\right) ^{3}.  \label{phi-prime}
\end{equation}
If $g^{\prime }=0$, Eq. (\ref{phi-prime}) exactly
coincides with Eq. (\ref{delta-phi}) for $m=1$, hence the function $\phi ^{\prime }(r)$ may be
identified as the corresponding zero mode. Of course, when $g$ is
a function of $x$ defined by Eq. (\ref{g}), which means $g^{\prime
}(x)=\delta (r-R)$, the term on the right-hand side of Eq.
(\ref{phi-prime}) does not allow $\phi ^{\prime }(x)$ to be the
zero mode; nevertheless, in the limit of $N\rightarrow
N_{\mathrm{Townes}}$, the soliton shrinks to a size much smaller
than $R$, hence $\left( \phi (R)\right) ^{3}$ becomes vanishingly
small, along with the above-mentioned term. Thus, in the limit of
$N=N_{\mathrm{Townes}}$, the function $\phi ^{\prime }(r)$
provides for a solution to Eq. (\ref{delta-phi}) with $m=1$,
making $N=N_{\mathrm{Townes}}$ the threshold of instability to the
azimuthal perturbations with $m=1$.

A notable feature of the stability diagram in Fig. 3 is that the
lower and upper stability borders meet and \emph{close down} the
stability region at $\rho =\rho _{\mathrm{\max }}\approx 0.95$,
which means that the nonlinearity-carrying annulus with the ratio
of the inner and outer radii exceeding the critical value, $\rho
_{\max }/R\approx \allowbreak 0.47$, cannot support stable
solitons. This conclusion implies that solitons cannot be stable
either in model (\ref{delta}) with the radial $\delta $-function.
Indeed, detailed consideration of that model reveals the region of
the azimuthal stability at $N>11.0$ and $\mu >-0.0116$, which
entirely belongs to the upper branch of the $\mu (N)$ curve in
Fig. 2(b), with $dN/d\mu >0$, i.e., the region is VK-unstable.

\section{Direct simulations}

To check the predictions for the stability of the solitons, and examine the
evolution of unstable ones, we have performed direct 2D simulations by dint
of the split-step Fourier method, employing a basis composed of $512\times
512$ modes. The size of the integration domain was $L\times L=60\times 60$,
with the center of the circle or annulus set at point $\left( x,y\right)
=(L/2,L/2)$, and the timestep $\Delta t=0.005$.

The simulations have confirmed the stability of the solitons in the region
between the lower and upper borders in Fig. 3, and instability outside of
this region. Figure 4(a) displays an example of the time evolution of $|\psi
(x,L/2)|$ (i.e., the profile of the cross section through the central point
along the $x$ axis) in a perturbed stable soliton, for $\rho =0$. On the
other hand, Figs. 4(b) and (c) demonstrate that (for the same case of $\rho
=0$) unstable solitons suffer collapse.
\begin{figure}[bp]
\includegraphics[height=3.5cm]{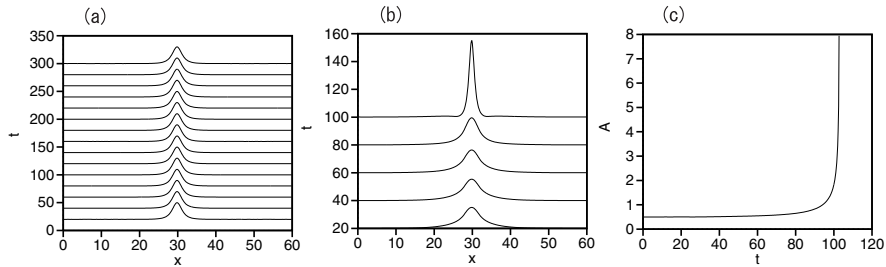}
\caption{(a) The evolution of $|\protect\psi (x,L/2)|$ (central
cross section) in a stable soliton for $\protect\rho =0$,
$\protect\mu =-0.189$ and $N=5.62$. (b) An example of collapse of
an unstable soliton, for $\protect\rho =0$, $\protect\mu =-0.0179$
and $N=6.175$. (c) The time dependence of the field amplitude,
i.e., maximum value of $|\protect\phi (x,y)|$, for the same case
as in (b).} \label{fig4}
\end{figure}

However, unstable solitons [ones belonging to the upper,
VK-unstable, part of the $\mu (N)$ curve in Fig. 1(b), with
$dN/d\mu >0$] whose norm is taken below a critical value,
$N_{\mathrm{cr}}^{(\mathrm{upper)}}\approx 5.99$ (for $\rho =0$),
which is \emph{higher} than the norm $N_{\mathrm{Townes}}\approx
5.85$ of the Townes soliton in the two-dimensional NLS\ equation,
neither collapse nor decay into radiation (in the NLS equation, a
pulse with $N<N_{\mathrm{Townes}}$ is bound to decay in the 2D
uniform space). Instead, the unstable soliton rearranges itself
into a \emph{stable breather}. Figures 5(a) and (b) display an
example of the evolution of breathers. In the simulations, the
breathers remain stable indefinitely long, their oscillations
getting more regular as $N$ decreases. The amplitude of the
oscillations, which we define as the root-mean square of the
variation of the soliton's amplitude, $A(t)\equiv |u(x=y=L/2,t)|$,
decreases with $N$, and it vanishes at another critical value,
$N_{\mathrm{cr}}^{(\mathrm{lower)}}\approx 5.449$. Up to the
numerical accuracy, the latter one is precisely the smallest value
of $N$ at which the stationary solitons exist for $\rho =0 $, see
Fig. 1(b). Thus, $N=N_{\mathrm{cr}}^{(\mathrm{lower)}}$ is not
only the point of the merger of the VK-stable and VK-unstable
branches of the solutions, but also the one at which the breathers
merge into the static solitons.
\begin{figure}[tbp]
\includegraphics[height=3.5cm]{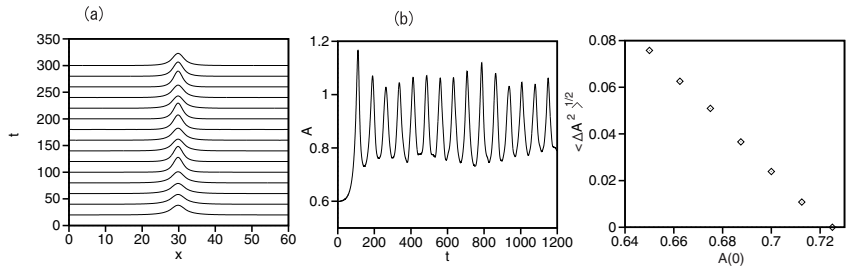}
\caption{(a) The time evolution of the cross-section profile,
$|\protect\psi (x,L/2)|$, of a breather, for $\protect\rho =0$,
$\protect\mu =-0.0399$, and $N=5.59$. The initial amplitude of the
soliton is $A(0)=0.6$. (b) Evolution of the amplitude of the
breathing soliton for the same case. (c) The amplitude of
intrinsic oscillations of the breather as a function of $A(0)$.}
\label{fig5}
\end{figure}

We stress that the existence of the stable \emph{axisymmetric} breathers up
to $N_{\mathrm{cr}}^{(\mathrm{upper)}}\approx 5.99$ does not contradict the
fact that the symmetry-breaking azimuthal instability occurs, for $\rho =0$,
at $N>N_{\mathrm{Townes}}\approx 5.85$, as explained above. Indeed, the
latter pertains to the angular instability of the static solitons, but not
breathers.

A noteworthy consequence of these results is the
\emph{bistability}: in the entire interval of values of the norm,
$N_{\mathrm{cr}}^{(\mathrm{lower)}}\approx
5.449<N<N_{\mathrm{Townes}}\approx 5.85$, where stable stationary
solitons are found (for $\rho =0$), they coexist with breathers.
On the other hand, in the adjacent interval,
$5.85<N<N_{\mathrm{cr}}^{(\mathrm{upper)}}\approx 5.99$, only
stable breathers are possible (and no stable objects exist for
$N>5.99$).

Stable breathers and bistability were found for $\rho >0$ as well. We note
that stable breathers were also found in the model based on the
one-dimensional GPE with a nonlinear OL [i.e., the nonlinearity coefficient
modulated in space as $\cos (kx)$] \cite{1d}. In the latter model,
bistability was observed too, as the breathers exist at the same values of
the norm at which stable stationary solitons are found.

As said above, all the solitons which are stable against collapse
in the model with $\rho =0$, are stable too against the azimuthal
perturbations. Actual instability against the azimuthal mode
(\ref{pert}) with $m=1$ occurs at $\rho >0$. To study the
azimuthal instability in direct simulations, we used an initial
condition in the form of a stationary soliton subjected to a weak
angular deformation. Figure 6 displays a typical example of the
development of the azimuthal instability for $\rho =0.55$. As a
result, the soliton does not split into fragments, which is a
generic result of the azimuthal instability of vortex-ring
solitons in uniform media \cite{ring}, but rather shifts from the
central point, $(x,y)=(30,30)$, to a position centered at
$(x,y)\approx (29,30)$. Because the norm of the soliton exceeds
$N_{\mathrm{Townes}}$, it then develops intrinsic collapse at the
new position, where the hole does not essentially affect its
dynamics. The shift of the soliton off the center and subsequent
collapse were found to be a generic outcome of the development of
the azimuthal instability. This feature can be easily explained by
the fact obvious in Fig. 3: all the solitons which are subject to
the azimuthal instability have $N>N_{\mathrm{Townes}}$, hence they
should collapse after being displaced away from the hole.
\begin{figure}[tbp]
\includegraphics[height=3.5cm]{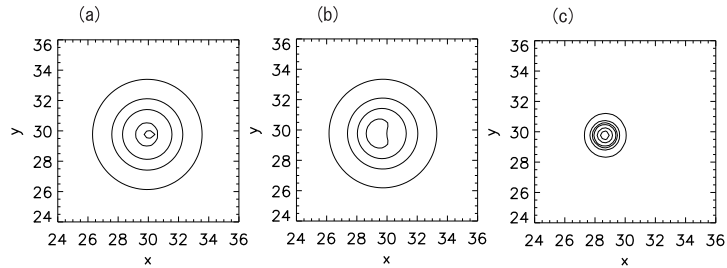}
\caption{Snapshots of contour maps of $|\protect\psi (x,y)|$ for
an azimuthally unstable soliton, taken at $t=5$ (a), $t=100$ (b)
and $t=122$ (c). In this case, $\protect\rho =0.55$, $\protect\mu
=-0.0782$ and $N=7.055$.} \label{fig6}
\end{figure}

As said above, direct simulations corroborate the stability of the
stationary solitons in the region between the two borders in Fig.~3. We
illustrate this conclusion in Fig. 7, which displays the time evolution of
the field amplitude (maximum value of $|\psi (x,y)|$) for $\rho =0.2$ and
three different values of the norm. The first soliton, with $N=5.702$,
belongs to the stability region in Fig. 3, and it is seen to be stable
indeed. Two other solitons, with $N=5.994$ and $N=6.01$, are azimuthally
unstable, which eventually leads to the collapse (after the spontaneous
off-center shift, as shown in Fig. 6). Note that, as $N=5.994$ is close to
the border of the azimuthal instability, the respective instability
development time is large.
\begin{figure}[tbp]
\includegraphics[height=3.5cm]{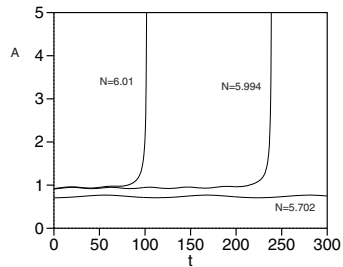}
\caption{The field amplitude vs. time, for weakly perturbed
solitons with $\protect\rho =0.2$ and (a) $N=5.702$, $\protect\mu
=-0.0758$, (b) $N=5.994$, $\protect\mu =-0.164$, and (c) $N=6.01$,
$\protect\mu =-0.169$.} \label{fig7}
\end{figure}

\section{Vortex solitons}

Besides the fundamental solitons considered above, Eq. (\ref{GPE}) also
gives rise to vortex solitons, in the form of $\psi =\phi _{S}(r)e^{-i\mu
t+iS\theta }$, with integer vorticity $S$ and real function $\phi (r)$
satisfying the equation [cf. Eq. (\ref{phi})]
\begin{equation}
\phi _{S}^{\prime \prime }+r^{-1}\phi _{S}^{\prime }-S^{2}r^{-2}\phi
_{S}+2g(r)\phi _{S}^{3}+2\mu \phi _{S}=0.  \label{phi-S}
\end{equation}
In particular, in the model with the radial $\delta $-function,
see Eq. (\ref{delta}), the vortex solution can be found in an exact form,
cf. Eqs. (\ref{Bessel}) and (\ref{A^2}):
\[
\phi _{S}(r)=A\left\{
\begin{array}{cc}
I_{S}\left( \sqrt{-2\mu }r\right) /I_{S}\left( \sqrt{-2\mu }R\right) , & r<R,
\\
K_{S}\left( \sqrt{-2\mu }r\right) /K_{S}\left( \sqrt{-2\mu
}R\right) , & r>R,\end{array}\right.
\]
\[
A^{2}=\frac{1}{2}\sqrt{-\frac{\mu }{2}}\left[ \frac{I_{S+1}\left(
x\right) +I_{S-1}(x)}{I_{S}\left( x\right) }+\frac{K_{S+1}\left(
x\right) +K_{S-1}(x)}{K_{S}\left( x\right) }\right]
|_{x=\sqrt{-2\mu }R}~.
\]
The norm of this solution can also be calculated in an analytical form.

An example of a vortex soliton, and the dependence $\mu (N)$ for these
solutions, are displayed in Figs. 8(a) and (b), for $\rho =0$ and $S=1$. The
figures show that a part of the solution family has $dN/d\mu <0$, hence it
is stable against radial perturbations, pursuant to the VK criterion.
\begin{figure}[tbp]
\includegraphics[height=3.5cm]{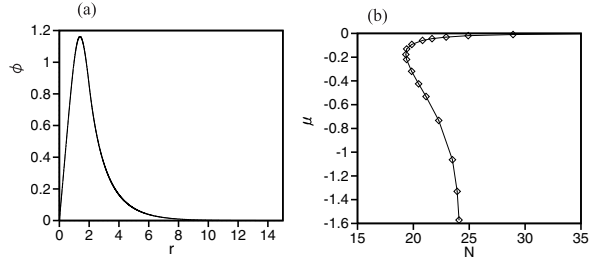}
\caption{(a) A typical example of profile $\protect\phi (r)$ for
the vortex soliton with $\protect\rho =0$, $S=1$, and
$N=23.9$,~$\protect\mu =-1.33$. (b) The $\protect\mu (N)$
dependence for the vortex-soliton family with $\protect\rho =0$
and $S=1$.}
\end{figure}

Comparing Fig. 8(b) to Fig. 1(b), one observes that the norm of
the vortices is much larger than the norm of the fundamental
solitons, which suggest that the vortex soliton may break up into
a set of fundamental ones (as said above, this is a typical
outcome of the development of azimuthal instability of vortex
solitons in uniform media \cite{ring}). Indeed, further analysis
demonstrates that the vortex solitons with $S=1$ are unstable
against azimuthal disturbances with $m=2$ [cf. Eq. (\ref{pert})].
An example, displayed in Fig. 9 for $\rho =0$, shows that the
instability splits the vortex into a set of two zero-vorticity
solitons, each then collapsing intrinsically, as its norm exceeds
the critical value, $N_{\mathrm{Townes}}\approx 5.85$. Before the
collapse, the soliton pair rotates in the counter-clockwise
direction. No example of a stable vortex soliton was found in the
model.
\begin{figure}[tbp]
\includegraphics[height=3.5cm]{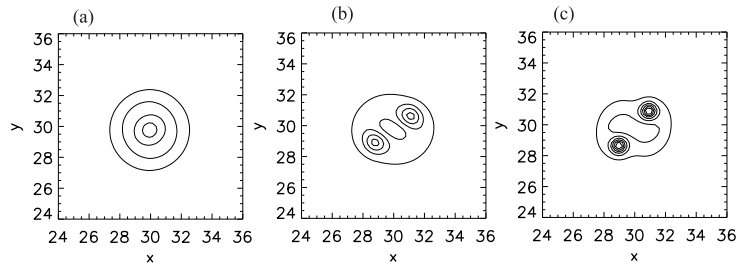}
\caption{Instability of vortex solitons is illustrated by a set of three
snapshots of the contour map of $|\protect\psi (x,y)|$ for a vortex with $S=1
$, $\protect\rho =0$, $N=23.9$ and $\protect\mu =-1.33$, taken at $t=50$
(a), $85$ (b), and $90$ (c).}
\label{fig9}
\end{figure}

\section{Conclusion}

The purpose of the work was to investigate the two-dimensional
Gross-Pitaevskii equation in which the attractive nonlinearity is limited to
a finite region in the form of a circle or annulus, including the case of a
narrow ring. In Bose-Einstein condensates trapped between a pair of
blue-detuned light sheets, this configuration can be implemented through the
Feshbach resonance by means of a properly configured magnetic or optical
field which controls the scattering length of collisions between atoms.
Using numerical and analytical methods, we have found a stability region for
axisymmetric fundamental (zero-vorticity) solitons in the model, which is
impossible in the case of the spatially uniform nonlinearity. It is
noteworthy that the stability borders of the solitons in the model with the
nonlinearity supported on the circle are completely determined by radial
perturbations, while in the annular model the upper stability border is set
by azimuthal modulations. The stability is limited to relatively broad
annuli, with the ratio of the inner and outer radii smaller than a critical
value, $\rho _{\max }/R\approx 0.47$. Moreover, the model gives rise to
bistability, as the stationary solitons coexist with stable axisymmetric
breathers. The stability region of the breathers extends, in terms of their
norm, to values exceeding the critical value corresponding to the Townes
soliton. The collapse threshold strongly increases with the radius of the
inner hole. Vortex solitons were constructed too, but they are unstable.
Essentially the same results were obtained also for a model in which,
outside of the circle or annulus, the nonlinearity is not zero but rather
repulsive (that case is not considered in the paper, as the configuration
with the zero nonlinearity is the most challenging one, as concerns the
stability of solitons). The results reported in this work suggest a
straightforward possibility to create stable two-dimensional matter-wave
solitons in Bose-Einstein condensates.

\section*{Acknowledgements}

We acknowledge valuable discussions with D. Frantzeskakis and P. G.
Kevrekidis. B. A. M. appreciates hospitality of the Department of Applied
Science for Electronics and Materials at the Interdisciplinary Graduate
School of Engineering Sciences, Kyushu University (Fukuoka, Japan). This
work was partly supported by the Grant-in-Aid for Scientific Research
No.17540358 from the Ministry of Education, Culture, Sports, Science and
Technology of Japan, and by the Israel Science Foundation through the
Center-of-Excellence in Research grant No. 8006/03.

\end{document}